\documentclass[twocolumn,prb,superscriptaddress,showpacs]{revtex4}

\def\xitil{\tilde\xi}
\def\betatilP{\tilde\beta P}
\def\betatil{\tilde\beta}
\def\gammatil{\tilde\gamma}
\def\mutil{\tilde\mu}
\def\dxi{\delta\xi}

  \usepackage[dvips]{graphicx}

\usepackage{dcolumn}
\usepackage{amsmath}
\usepackage{latexsym}

\begin{document}

\title[Period-doubling-bifurcation readout]{Period-doubling-bifurcation readout for a Josephson qubit}
\author{Alexander~B.~Zorin}
\affiliation{Physikalisch-Technische Bundesanstalt, Bundesallee 100, D-38116 Braunschweig, Germany}
\affiliation {Skobeltsyn Institute of Nuclear Physics, Moscow State University, 119899 Moscow, Russia}
\author{Yuriy~Makhlin}
\affiliation{Landau Institute for Theoretical Physics, Kosygin st. 2, 119334, Moscow, Russia}
\affiliation{Moscow Institute of Physics and Technology, 141700, Dolgoprudny, Russia}


\begin{abstract}
We propose a threshold detector with an operation principle, based
on a parametric period-doubling bifurcation in an externally pumped
nonlinear resonance circuit. The ac-driven resonance circuit
includes a dc-current-biased Josephson junction ensuring parametric
frequency conversion (period-doubling bifurcation) due to its
quadratic nonlinearity. A sharp onset of oscillations at
the half-frequency of the drive allows for detection of small
variations of an effective inductance and, therefore, the read-out of the quantum
state of a coupled Josephson qubit. The bifurcation
characteristics of this circuit are compared with those of the
conventional Josephson bifurcation amplifier, and its possible advantages are discussed.


\end{abstract}
\maketitle

The problem of an efficient readout of solid state quantum systems
including Josephson qubits (see, e.g., Ref.~\onlinecite{Makhlin}) is of
high importance from both theoretical and practical points of view.
The dispersive readout techniques based on the radio-frequency
measurement of reactive electrical parameters (for example, the
Josephson inductance~\cite{Z-Phys-C-JETP} or quantum Bloch
capacitance~\cite{Sillanpaa,Duty}) received significant recognition, since they
allow one to minimize the backaction of the readout circuit on a Josephson
qubit. Recently, particular interest was focused on such systems
operating in the non-linear resonance regime (Duffing oscillator),
which was possible due to a cubic non-linearity of the supercurrent in a
zero-phase biased Josephson junction~\cite{Ithier,Lupascu,Siddiqi-2006,Vijay-2009}. In
this regime, under the action of a weak signal and/or fluctuations the
circuit undergoes a bifurcation, i.e., a transition between two
stable oscillatory states~\cite{Dykman-80}. The successful idea of
application of such a Josephson Bifurcation Amplifier (JBA) for
measurements of a qubit was first proposed by Siddiqi et al.~\cite{Siddiqi-2004},
and this has served for us as a motivation for the
development of a readout based on another type of bifurcation in
superconducting non-linear circuits.
\begin{figure}[b]
\begin{center}
\includegraphics[width=3.4in]{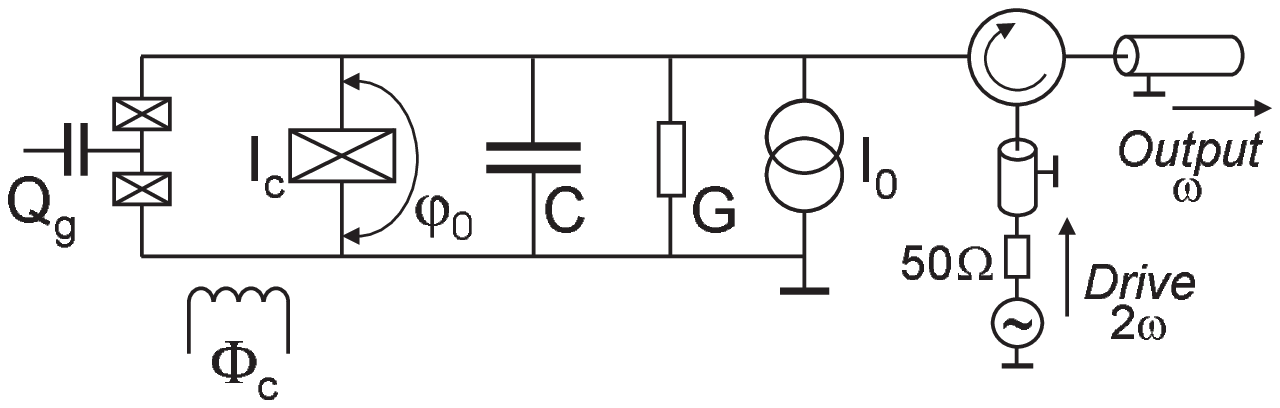}
\caption{Electric circuit diagram of the period-doubling bifurcation detector
with microwave-based readout. The resonator is formed by the
inductance of a non-linear Josephson junction (large crossed box),
biased at a non-zero phase value $\varphi_0$, and the capacitance $C$.
\\
The linear losses are accounted for by the conductance $G$.
The resonator is coupled to a charge-phase qubit formed by a superconducting
single electron transistor with capacitive gate (left) and attached to the
Josephson junction. The qubit operation
at the optimal point for an arbitrary bias $I_0$ is ensured by a
proper value of the
external magnetic control flux $\Phi_c$, applied to the qubit loop,
and the gate charge $Q_g$ on the qubit island.} \label{fig:EqvSchm}
\end{center}
\end{figure}

In this paper we propose a readout circuit, whose operation principle
is based on excitation of half-harmonic oscillations, i.e., a Period
Doubling Bifurcation (PDB), occurring in an rf-driven resonance
circuit with quadratic non-linearity of reactance. The essential
difference of the PDB from the JBA regime consists in the parametric
nature of the PDB resonance manifesting itself in abrupt switching
from zero-oscillation state into the dynamic state with a double
period and appreciable amplitude of the oscillations~\cite{Migulin}.
This regime may be favorable for an output-stage preamplifier
receiving in the case of the PDB a signal with zero background.
Moreover, as we shall show below, the switching characteristics of
our circuits are somewhat different from those of conventional JBA;
in particular, we find that in addition to better contrast between
two possible stationary states of the PDBA, it may have a narrower switching region.

The PDB circuit (see Fig.~\ref{fig:EqvSchm}) comprises a dc-current-biased Josephson
junction with the critical current $I_c$, capacitance $C$ including
the self-capacitance of the junction with, possibly, a contribution of an
external capacitance, the linear shunting conductance $G$, as well as an
attached qubit, presented here as a charge-phase qubit~\cite{Vion,Z-Phys-C-JETP}. The circuit
is driven by a harmonic signal $I_{\textrm{ac}}=I_A \cos 2\omega t$ at a
frequency close to the double frequency of
small-amplitude plasma oscillations $\omega_p$, i.e., $\omega \approx
\omega_p$.

Neglecting fluctuations, the dynamics of the bare system (excluding
the qubit, whose quantum state only slightly changes the plasma
frequency of the entire circuit, $\omega_p \rightarrow
\widetilde{\omega}_p$) is governed by the model of a resistively shunted
junction~\cite{RSJ}:
\begin{equation}
\label{circuit-Eq} \frac{\hbar C}{2e} \frac{d^{2}\varphi}{dt^{2}}
+ \frac{\hbar G}{2e} \frac{d\varphi}{dt}
+ I_c\:\sin\varphi = I_0+I_{\textrm{ac}}\,,
\end{equation}
where the finite current bias $I_0<I_c$ ensures a dc phase drop
$\varphi_0 = \arcsin(I_0/I_c)$ across the Josephson junction. The
small--ac-signal expansion ($x \ll 1$) of the Josephson supercurrent
term includes the following components: $\sin\varphi =
\sin(\varphi_0+x) \approx \sin\varphi_0
(1-x^2/2+x^4/24)+\cos\varphi_0 (x-x^3/6)$. The angular frequency of
small oscillations of $\varphi$ around $\varphi_0$ is $\omega_p =
 (\cos\varphi_0)^{1/2}\omega_{p0}$, where the bare plasma frequency
is $\omega_{p0} = (2eI_c/ \hbar C)^{1/2}$.

\begin{figure}[b]
\begin{center}
\includegraphics[width=3.3in]{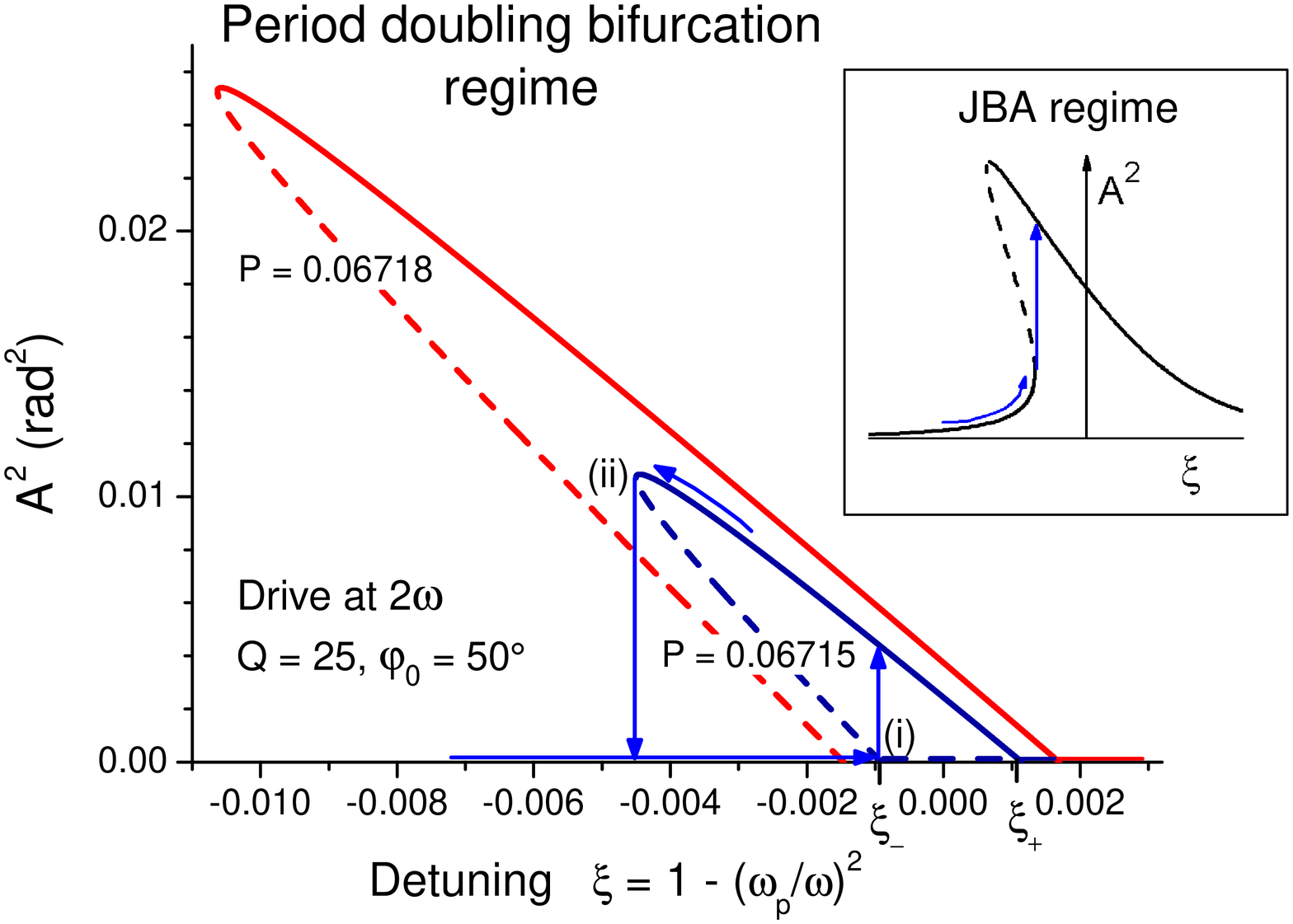}
\caption{(Color online) Intensity $A^2$ of oscillations of
the Josephson phase at frequency $\omega$ versus frequency detuning $\xi$ for two
amplitudes of the pumping signal (at frequency $2\omega$). Dashed
lines show unstable states. When the detuning approaches a bifurcation point ((i) or (ii)), PDB occurs
(vertical arrows from or to the zero state, respectively).
For comparison, a typical resonance curve of a JBA is sketched in the inset.} \label{fig:res-curve}
\end{center}
\end{figure}

Using the dot to denote derivatives with respect to the
dimensionless time $\tau = \omega t$, we write the equation of motion for $x$ in the form:
\begin{equation}
\label{normalized_Eq} \ddot{x} + x = \xi x -2\theta \dot{x} + \beta
x^2 + \gamma x^3 -\mu x^4 +3P\,\cos2\tau.
\end{equation}
The dimensionless coefficients in this equation are
\begin{equation}
\xi = 1-\kappa \ll 1,\qquad \theta=G/2\omega C \equiv 1/2Q \ll
1,\label{coeff:1}
\end{equation}
\begin{equation}
\beta=12\mu=(\kappa \tan\varphi_0) /2, \quad  \gamma = \kappa/6,\quad
3P = \kappa I_A/I_c,\label{coeff:2}
\end{equation}
where $\kappa= (\omega_p/\omega)^2\approx1$ and $Q$ is the quality factor.
The quadratic non-linear
term ($\propto \beta$) in Eq.~(\ref{normalized_Eq}) ensures
parametric down-conversion from the drive frequency
$2\omega$. Note, that similar frequency conversion, and therefore
the PDB effect, is also possible without quadratic non-linearity, if instead
of the driving force $3P\cos2\tau$ in Eq.~(\ref{normalized_Eq}) the
circuit is parametrically driven by a term of the form $xP'\cos2\tau$
(see, for example, Refs.~\onlinecite{Migulin,Dykman-98}). This case
can be realized by a periodic
modulation of the critical current of the zero dc-biased Josephson
element $I_c$ in a split dc-SQUID configuration by using an alternating
magnetic flux driving in the SQUID loop~\cite{Delsing}.

The leading terms in the solution of Eq.~(\ref{normalized_Eq}) have the form $x \equiv
y - P \cos 2\tau$, where $y(\tau)$ denotes oscillations at the frequency
$\approx\omega$, and the second term is the forced oscillation at the
drive frequency $2\omega$; $x(\tau)$ also contains other harmonics at multiple frequencies, which strongly influence its dynamics and stationary states~\cite{Shteinas}, cf.~Eqs.~(\ref{eq:xibegatil}--\ref{eq:mutil}) below.
We apply the method of slowly-varying amplitudes by introducing slow variables~\cite{Migulin},
\begin{equation}
\label{A-and-alpha} y= A \cos (\tau - \alpha)\,,
\qquad \dot{y}\approx -A \sin (\tau - \alpha)\,.
\end{equation}
The variables $A(\tau)$ and $\alpha(\tau)$ present the amplitude and phase (relative to the drive) of the oscillation at the half-frequency
of the drive; they vary weakly over the period of these oscillations (with dimensionless rates $\ll1$). Accordingly,
\begin{equation}
u=A\cos\alpha\,,\qquad v=A\sin\alpha\,
\label{eq:uv}
\end{equation}
are two quadratures of these oscillations, $u^2+v^2=A^2$. The dynamics of the slow variables is governed by the equations:
\begin{equation}
\label{dot-A-alpha}
\left(\begin{array}{c}\dot{A}\\A\dot\alpha\end{array}\right)
= - \int_\tau^{\tau+2\pi} \frac{d\tau}{2\pi}
\varepsilon F(x,\dot{x},\tau)
\left(\begin{array}{c}\sin(\tau-\alpha)\\\cos(\tau-\alpha)\end{array}\right)
\,,
\end{equation}
with averaging over a $2\pi$-period of the oscillations at frequency $\omega$ ($1$ in dimensionless units),
where the function in the integrand
\begin{equation}
\label{eF} \varepsilon F(x,\dot{x},\tau) = \xi x -2\theta \dot{x} +
\beta x^2 + \gamma x^3 -\mu x^4
\end{equation}
includes small terms at frequency $\omega$ and large terms at the
drive frequency $2\omega$ and its higher harmonics. The averaging
over the period of oscillations in Eq.~(\ref{dot-A-alpha}) yields a pair of reduced equations for the amplitude and phase:
\begin{eqnarray}
\dot A &=& - \theta A - \frac{1}{2}A\sin2\alpha (\betatilP-\mutil PA^2) \,, \label{dot-A}\\
\dot\alpha &=& \frac{\xitil}{2}-\frac{\betatilP}{2}\cos2\alpha + \frac{3}{8}\gammatil A^2 +\mutil PA^2\cos2\alpha \label{dot-alpha}\,.
\end{eqnarray}
The coefficients $\xitil(P)$, $\betatil(P)$, $\gammatil(P)$, $\mutil(P)$ to the leading order in $P^2$ are given by~\cite{Shteinas}
\begin{eqnarray}
\xitil = \xi,&\qquad& \tilde\beta = \beta,\qquad
\gammatil = \gamma + \frac{10}{9}\beta^2,\label{eq:xibegatil}\\
\mutil &=& \mu -\frac{15}{16}\beta\gamma+\frac{7}{24}\beta^3\,,\label{eq:mutil}
\end{eqnarray}
which implies that $\gammatil\approx (3+5\tan^2\varphi_0)/18$ and $\mutil \approx(7/192)\tan\varphi_0(\tan^2\varphi_0-1)$.
Corrections of order $P^2$ to these coefficients do not change further analysis qualitatively,
but only slightly modify the results quantitatively.

Equation~(\ref{dot-A}) always has a trivial solution $A=0$.
In the limit of weak pumping ($P \ll 1$)
and small resulting oscillations ($A^2 \ll 1$), the last terms ($\propto \mutil P$) on
the right hand side of Eqs.~(\ref{dot-A}--\ref{dot-alpha}) can
be neglected, and the oscillation amplitude of the non-zero stationary solutions ($\dot A=\dot\alpha=0$, $A \neq 0$) may be
found explicitly~\cite{Migulin}:
\begin{equation}
\label{A-stat} A_{\pm}^2 = \frac{4}{3\gammatil}\left[-\xitil \pm \sqrt{(\betatilP)^2 -4 \theta^2}\right].
\end{equation}
For the pumping amplitude exceeding the threshold set by dissipation, $|\betatilP| > 2\theta$, the values
\begin{equation}
\label{ksi-pm}\xitil_{\pm}= \pm \sqrt{(\betatilP)^2 -4 \theta^2}
\end{equation}
yield the range of frequency detunings, $\xi_- < \xi< \xi_+$, within
which the zero solution is unstable. In this range the system switches into the
oscillating state with a finite amplitude $A_+$ given by
Eq.~(\ref{A-stat}). For $\xi<\xi_-$ the parametric resonance curve
is multivalued with the stable trivial $A=0$ and nontrivial $A_+$
solutions, while the solution $A_-$ is unstable.
Taking into account higher (e.g., $\propto\mutil$) terms in Eqs.~(\ref{dot-A}--\ref{dot-alpha}) ensures
that $A_+(\xi)$ and $A_-(\xi)$ merge, limiting both the amplitude
$A_+$ and the the range of bistability in $\xi$;
for stronger drive even higher nonlinearities become important.
The shape of the resonance curve,
calculated numerically from Eqs.~(\ref{dot-A}--\ref{dot-alpha}),
is shown in Fig.~\ref{fig:res-curve} for several values of the drive amplitude
$3P$ just above the excitation threshold.
However, for further considerations of the threshold behavior the higher nonlinearities are not crucial, and below we neglect the $\mutil$-terms.

\begin{figure}
\begin{center}
\includegraphics[width=\columnwidth]{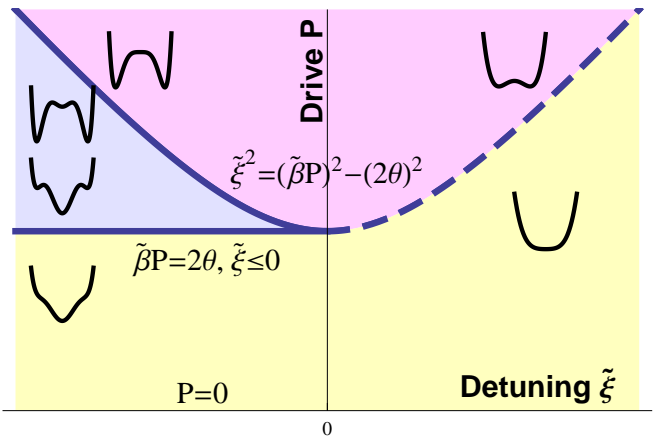}
\caption{(Color online) Stability diagram of the period-doubling bifurcation readout.
Stationary oscillatory solutions appear and disappear as the detuning $\xi$ and driving amplitude $P$ are varied.
This is indicated by the sketched potential curves, which show stable (minima) and unstable (maxima) states. The former include:
only the zero solution in the lower region, only a pair of equal-amplitude solutions $A_+$, $A_+^*$ with a $\pi$ phase shift in the upper region, and the zero and the pair $A_+$, $A_+^*$ in the left, `bistable' sector-shaped region. The equations in the figure describe the bifurcation boundaries.} \label{fig:stabDiag}
\end{center}
\end{figure}

The stability diagram of the system in the space of the control parameters, the detuning $\tilde{\xi}$ and the driving amplitude $P$ is shown in Fig.~\ref{fig:stabDiag}. The parameter plane is divided in three regions, with the following stable-state amplitudes (cf.~Eq.~(\ref{A-stat})):
$A=0$ in the lower region, $A_+$ in the upper region, and both $0$ and $A_+$ in the `bistable' sector (this region is limited by two solid
lines).
The bifurcation lines are given by the relations $A_+=A_-$ (lower left horizontal solid line), $A_-=0$ (i.e., $\xi=\xi_-(P)$, upper solid curve)
and $A_+=0$ (i.e., $\xi=\xi_+(P)$, dashed curve).
The coordinates of the triple point are $\xi=0$ and $P=2\theta/\betatil\approx 2/(Q\tan\varphi_0)$.

\begin{figure}[b]
\begin{center}
\includegraphics[width=2.6in]{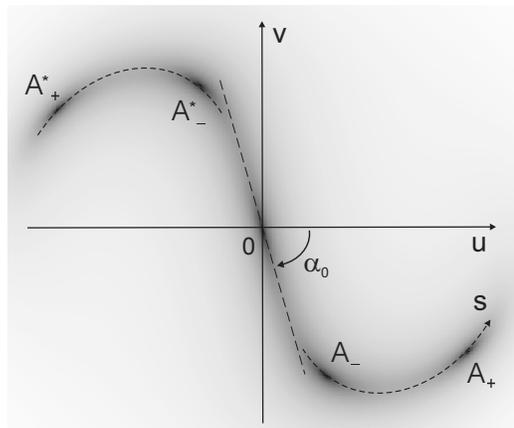}
\caption{Contour velocity plot calculated from Eqs.~(\ref{dot-A}--\ref{dot-alpha}) and (\ref{dot-u-v})
for the parameters corresponding to a multivalued solution.
The absolute value of velocity $|\textbf{v}|$ is lower in darker areas.
The curvilinear trajectory along the valley of minimal
velocity (short-dash line) is parametrized by $s$. The straight dashed
line at origin indicates the most probable direction of escape from the zero
state close to the bifurcation.} \label{velocity-plot}
\end{center}
\end{figure}

Equations for the quadrature components of the velocity field,
\begin{equation}
\label{dot-u-v}
\textbf{v} =\left(\begin{array}{c}\dot{u}\\\dot{u}\end{array}\right)
=
\left(\begin{array}{c} \dot{A}\cos\alpha-A\dot{\alpha}\sin\alpha\\\dot{A}\sin\alpha+A\dot{\alpha}\cos\alpha\end{array}\right),
\end{equation}
where $\dot{A}$ and $\dot{\alpha}$ are given by (\ref{dot-A}--\ref{dot-alpha}),
can be represented as Hamiltonian equations of motion with friction:
\begin{equation}
\label{dUuv} \dot{u}= -\partial_v H -\theta u ,
\quad\dot{v}= \partial_u H -\theta v\,,
\end{equation}
or, equivalently,
\begin{equation}
\label{dUAalpha} \dot A = -\theta A - \frac{1}{A}\partial_\alpha H,
\quad A\dot\alpha = \partial_A H \,,
\end{equation}
where the Hamiltonian is given by:
\begin{equation}
H = (\xitil - \betatilP \cos2\alpha) \frac{A^2}{4}
+
\frac{3}{32}\gammatil
A^4.
\end{equation}
This Hamiltonian for the slow variables can be obtained from the Hamiltonian
for the physical quantities~\cite{Shteinas}.
Figure~\ref{velocity-plot} shows a contour plot of the absolute value
of the velocity $|\textbf{v}|= (\dot{u}^2 + \dot{v}^2)^{1/2} =(\dot A^2+A^2\dot\alpha^2)^{1/2}$
in the case of a multivalued stationary solution. One
can see the darker $S$-shaped narrow valley, where the motion is slow along
the curvilinear $s$-axis.
The black spots in this area show the stationary solutions,
which are the stable focus at zero, $A=0$, the stable foci $A_+$ and $A^*_+$ corresponding to
equal-amplitude oscillations with a mutual phase shift of $\pi$, and the unstable saddles $A_-$ and $A^*_-$
(also with a mutual $\pi$-shift).
For weak dissipation these `saddle points' are the lower points of the barriers separating the
basins of attraction of the foci in the landscape of $H$. Thus the most probable escape path from the zero state is along
the $S$-shaped valley.

\begin{figure}
\begin{center}
\includegraphics[width=2.8in]{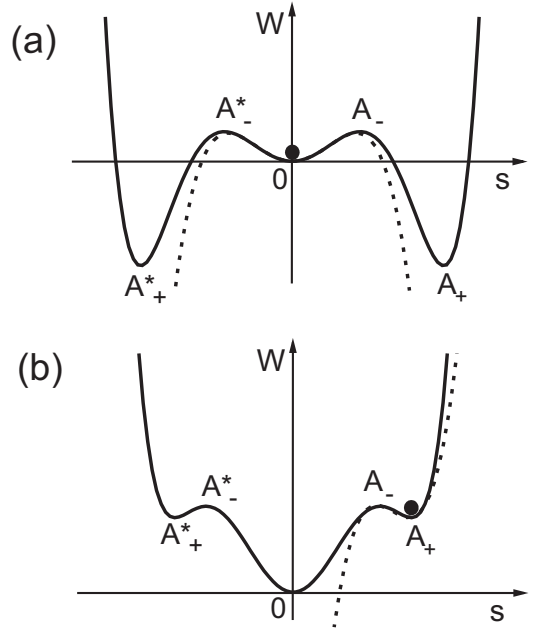}
\caption{A sketch of the potential $W(s)$ along the $S$-shaped region
(see Fig.~\ref{velocity-plot}) for two ultimate cases:
(a) bifurcation $0 \rightarrow A_+ \,(A^*_+)$ (near the upper solid line in Fig.~\ref{fig:stabDiag}) and (b) bifurcation $A_+ \rightarrow 0$ (near the lower solid line in Fig.~\ref{fig:stabDiag}).
Near the bifurcations the dashed lines approximate the energy barriers
by a quartic and cubic polynomials, respectively.} \label{fig:bifurc-W(s)}
\end{center}
\end{figure}

In the vicinity of the bifurcation point $\xi_-$ within the bistable region, the height of the energy barriers is small,
and one can show that there is a separation of time scales, which can be used to solve the dynamics:
the fast relaxation from outside towards the $S$-shaped valley is followed by slow dynamics along the valley.
In this region the points $A_-$ and $A^*_-$ are close to the origin, $u = v =0$, and the slope $\alpha_0$ of
the valley at the origin can be found from Eq.~(\ref{dot-A}):
\begin{equation}
\label{tan-alpha} \sin 2\alpha_0 = -\eta^{-1}, \textrm{\quad where \quad} \eta = \betatilP/2\theta.
\end{equation}
To describe the slow motion along the valley near the origin, where one can
use the amplitude $A$ as a coordinate, we first solve an equation for the fast motion
in the axial $\alpha$-direction (variable $\alpha$ relaxes fast, with a typical rate of $\theta$).
To find the subleading nonlinear terms in the equation of motion along the valley,
one needs to take into account the deviation of the valley near the origin from a straight line.
The resulting equation of motion can be represented in the form of an easily solvable 1D equation (cf.~Ref.~\onlinecite{Dykman-80}):
\begin{equation}
\label{dot-s} \frac{ds}{d \tau} = -\frac{dW(s)}{ds}\,,
\end{equation}
where the pseudopotential is to the lowest orders a biquadratic
polynomial, $W=as^2-bs^4$, where
\begin{equation}
\label{coeff-ab}
a = -\frac{\delta\xitil}{4}\sqrt{\eta^2-1} \,,\qquad
b = \frac{3}{32} \gammatil\sqrt{\eta^2 - 1}
\,.
\end{equation}
For $\xi<\xi_-$ ($\dxi<0$) one finds that $a>0$.
Thus, when $\xi$ crosses $\xi_-$ from above, the zero unstable stationary
solution bifurcates and separates into a stable solution at zero and two
symmetric unstable solutions $A_-$ (Fig.~\ref{fig:bifurc-W(s)}).
This property makes it sensitive to small changes of the circuit parameters
(in particular, to the qubit state via its effective Josephson inductance, which modifies the detuning $\xi$).
The switching characteristics of such a detector can be found from
the analysis of this system in the presence of noise, which results in
a finite width of the transition. To describe the bifurcation-based readout, one needs to find
the tunneling rate out of the shallow well $W(s)$ near the bifurcation.

Small fluctuations due to the conductance $G$ are taken into account by adding a
noise term $\delta I$, with the 
spectral density $S_{I}(\omega) = 2\hbar\omega G \coth\frac{\hbar\omega}{2k_BT}$,
to the rhs of Eq.~(\ref{circuit-Eq}).
This gives rise to independent fluctuations of the two quadratures.
Their correlation functions are
\begin{equation}
\langle \delta u(\tau)\delta u(\tau') \rangle =
\langle \delta v(\tau)\delta v(\tau') \rangle = 2 T_{\rm eff} \delta(\tau-\tau'),
\end{equation}
with $\langle \delta u(\tau)\delta v(\tau') \rangle =0$ or
\begin{equation}
\langle \delta A(\tau)\delta A(\tau')\rangle
= A^2 \langle \delta \alpha(\tau)\delta \alpha(\tau')\rangle =
2 T_{\rm eff} \delta(\tau-\tau'),
\end{equation}
with $\langle \delta A(\tau)\delta\alpha(\tau')\rangle = 0$, where the effective temperature
\begin{equation}
T_{\rm eff}= \frac{ \kappa^2 S_I(\omega)\omega}{8 (I_c \cos\varphi_0)^2}
\stackrel{\mathrm{low-}\omega}{\longrightarrow}
\frac{\kappa^2 k_BTG\omega}{2(I_c\cos\varphi_0)^2}\,,
\label{eq:Teff}
\end{equation}
and the latter expression holds in the low-frequency (classical) limit $\hbar\omega\ll k_BT$.
Upon the reduction to the 1D equation (\ref{dot-s}) this gives fluctuations with the same noise power,
\begin{equation}
\langle \delta s(\tau)\delta s(\tau')\rangle =
2 T_{\rm eff} \delta(\tau-\tau') \,,
\end{equation}
which affect the motion along the $s$-coordinate.

Adding the Langevin term $\delta s(\tau)$ on the right-hand side of Eq.~(\ref{dot-s}), one can derive and then solve a 1D Fokker-Planck
equation~\cite{Dykman-80} (in fact, a Smoluchowski equation since the `mass' term, $\propto d^2s/d\tau^2$, is absent in Eq.~(\ref{dot-s}))
for the probability density $w(s,t)$ to find the system at the point $s$ at the time $t$,
\begin{equation}
\label{FP-eq} \frac{\partial w}{\partial t} = \frac{\partial}{\partial s}\left[ \frac{\partial W(s)}{\partial s}w \right]
+ T_{\rm eff} \frac{\partial^2 w}{\partial s^2}\,.
\end{equation}

The escape rate out of the zero metastable state $A=0$ to the stable
state $A_+$ or $A^*_+$ is given by the Kramers formula~\cite{Kramers}
reflecting the activational behavior of the system,
\begin{equation}
\Gamma = 2 (\omega_A/2\pi) e^{-\Delta W/T_{\rm eff}} \,,
\end{equation}
where the factor $2$ accounts for two escape possibilities (to the left or right wells).
For the overdamped case of a zero-mass particle the formula for $\omega_A$ is given, for example, in
Ref.~\onlinecite{Melnikov}. The prefactor $\omega_A$ is determined by the geometrical
mean of the curvatures of $W(s)$ at the bottom of the central well (equal to $2a$) and at the top of the barrier (equal to $4a$), i.e.,
\begin{equation}
\label{prefactor}
\omega_A = 2\sqrt{2} a \omega
= \omega \frac{|\dxi|}{\sqrt{2}} \sqrt{\eta^2-1} \,.
\end{equation}
The barrier height to the lowest order in $\dxi=\xi-\xi_-$ is
\begin{equation}
\Delta W = a^2/4b \propto (\xi-\xi_-)^2\,.
\label{eq:barrier}
\end{equation}

\begin{figure}
\begin{center}
\includegraphics[width=\columnwidth]{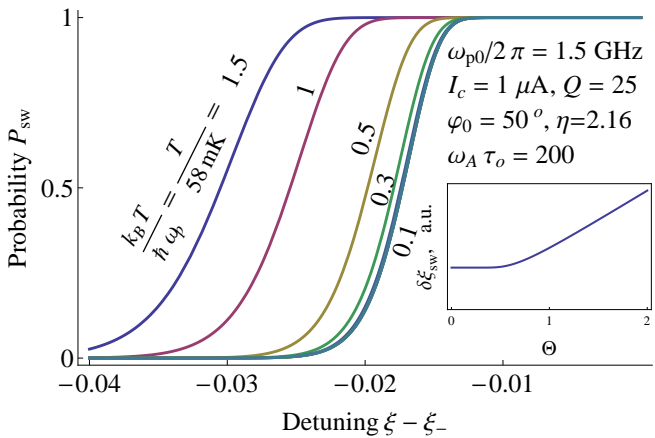}
\caption{(Color online) Switching curves for the period-doubling bifurcation readout at various temperatures.
Note saturation at low $T$, governed by `quantum noise' at $k_BT\lesssim\hbar\omega$.
{\it Inset}: low-$T$ saturation, illustrated by the curves' width $\delta\xi_\mathrm{sw}$ (inverse slope at $P=0.5$, $\propto\sqrt{S(\omega)}$) as a function of $\Theta \equiv \sqrt{k_BT/\hbar\omega}$.
}
\label{fig:switch}
\end{center}
\end{figure}

Typical switching curves (switching probability $P_\mathrm{sw}=1-e^{-\Gamma\tau_o}$ during some observation time $\tau_o$ vs. $\xi$) are shown in Fig.~\ref{fig:switch} for various temperatures for a set of typical circuit parameters. Note that the position and the width of the switching curve (see inset) saturates at low temperatures. This effect is not a manifestation of the real quantum tunneling, but is rather linked to the fact that activation in the rotating frame of the first harmonic Eqs.~(\ref{A-and-alpha},\ref{eq:uv}), i.e., the low-frequency noise in that frame, is given in the laboratory frame by the noise at a finite frequency $\omega$, cf.~Eq.(\ref{eq:Teff}) and above.

Equation~(\ref{eq:barrier}) implies that the width (along the detuning axis $\xi$) of the switching curves,
given by the inverse slope $(dP/d\xi)^{-1}$ at $P = 0.5$,
scales as 
$\delta\xi_\mathrm{sw}\propto \sqrt{T}$
above saturation. Thus as $T \rightarrow 0$, it falls off slightly slower than that for the `standard' Josephson bifurcation amplifier~\cite{Siddiqi-2004,Siddiqi-MQC2004}, where $\delta\xi_{\rm sw}\propto T^{2/3}$. This (minor) difference stems from  symmetry of the PDBA w.r.t. a shift by a drive period: $\alpha\to\alpha+\pi$. This symmetry implies that the generic form of the 1D potential $W$ near the bifurcation is
$\varepsilon s^2-as^4$ unlike $\varepsilon s -as^3$ for the JBA. Here $\varepsilon$ measures the distance from the bifurcation, and $s$ is the relevant coordinate in phase space.
However, this symmetry can be broken, and the stronger effect of cooling ($\delta\xi_{\rm sw}\propto T^{2/3}$) restored, by a weak admixture at frequency  $\omega$ to the drive signal. An alternative strategy consists in using another bifurcation point, where $A_+=A_-$ in Fig.~\ref{fig:res-curve} (on lowering the detuning $\xi$, the system follows the solution $A_+$ until it merges with $A_-$, where it switches abruptly to zero; there is no symmetry around this point).

Thus we have suggested two protocols of operation of the PDBA (with potentials shown in Fig.~\ref{fig:bifurc-W(s)} and operation indicated by arrows in the parametric-resonance plot Fig.~\ref{fig:res-curve}): one of them involves switching from the zero state to a large-amplitude stable state $A_+$ near the bifurcation point $\xi_-$, and the other involves a reverse switching from the large-amplitude state $A_+$ to zero near the merging point of $A_+$ and $A_-$. Note that in both cases to perform a read-out, that is to find out if a switching has occurred, one needs to distinguish a zero state from a large-amplitude state. This should be contrasted with the JBA, where two finite-amplitude (and often, similar-amplitude, but different-phase) states have to be distinguished. From this viewpoint, the PDBA may be more convenient in practical applications. Other protocols can also be discussed (cf.~Ref.~\onlinecite{Dykman08}).

The readout of a coupled qubit is based on the shift in the plasma frequency (and thus of the switching curve) due to different Josephson inductances in two qubit states. The inductance values depend on the type of qubit and its parameters. During the readout the qubit state is encoded in the resulting oscillations of the PDBA by tuning the control parameters (such as the drive frequency and amplitude, i.e., $\xi$ and $P$) to a point with the maximal difference (contrast) between the two switching curves. 
High contrast is reached when the shift in the plasma frequency exceeds the width of the switching curve. 
In an ideal arrangement, this contrast reaches 100\%: $P_{\rm sw}=0$ and $1$ for two qubit states.
For the PDBA the contrast reaches values comparable to those for the JBA with similar circuit parameters (for example, about 0.3\% in frequency sensitivity for the parameters of Fig.~\ref{fig:switch} at low $T$, that is sufficient for reliable readout of the charge-phase qubit ~\cite{Ithier} shown in Fig.~\ref{fig:EqvSchm}). 
Further optimization of the PDBA parameters is possible.

Let us compare the switching curves for the PDBA and JBA~\cite{Ithier} near the upper critical lines of the bistability region ($\eta_\uparrow$ in the notations of Ref.~\onlinecite{Ithier}). We consider the tunneling exponents as functions of the dimensionless deviation of the drive amplitude
from the bifurcation, $1-P/P_-$, for the PDBA, and we use the same notation, instead of $1-\eta/\eta_\uparrow$, for the JBA.
According to Refs.~\onlinecite{Siddiqi-MQC2004,Ithier}, for the JBA
\begin{equation}
\left|1-\frac{P}{P_-}\right| \sim \left( \frac{k_\mathrm{B}T}{E_J|\xi|} \right)^{2/3}.
\end{equation}
For the PDBA, we replace the difference $(\xi-\xi_-)$ in Eq.~(\ref{eq:barrier}) by $\partial_P\xi_-(P_--P)$.
Assuming that the detuning $\xi\gtrsim1$ (that is $\delta\omega\gtrsim\omega/Q$), we find that the tunneling exponent is of order
\begin{equation}
\frac{\Delta W}{T_{\rm eff}}\approx
Q^2 \frac{6\gammatil\cos\varphi_0 E_J}{k_\mathrm{B}T} |\xi|^3
\left(1-\frac{P}{P_-}\right)^2.
\end{equation}
Thus increasing the detuning (and the corresponding driving amplitude $P_-(\xi)$) can suppress
the width of the relevant switching curve (switching probability-vs.-drive amplitude $P$):
\begin{equation}
\left|1-\frac{P}{P_-}\right| \sim \left( \frac{k_\mathrm{B}T}{E_J|\xi|} \frac{1}{Q^2|\xi|^2} \frac{1}{6\gammatil\cos\varphi_0} \right)^{1/2}.
\end{equation}


We note that various operation protocols of the readout device based on PDBA are possible, and one can force
crossing the bifurcation region and switching between the oscillating states by tuning various parameters.
In particular, the current bias $I_0$, the amplitude and frequency of the drive can be used for engineering
a metapotential of desired shape and, therefore, optimization of the readout.
In our analysis we have focused on the noise-induced activation over the barrier in this metapotential.
As we saw (cf. above Eq.~(\ref{eq:Teff})), the effective `temperature' is set by the noise
level at frequency $\omega$ and saturates on lowering the temperature $T$ below $\hbar\omega/k_B$.
This low-$T$ regime may also be thought of as `quantum activation'~\cite{MarDyk06,Dykman08}.
One could also consider the quantum tunneling~\cite{DmiDyak}. However, in similar systems the corresponding
tunneling rate is exponentially small, especially close to the bifurcation point (cf.~Refs.~\onlinecite{Dykman08,MarDyk06}).

In conclusion, we have suggested to use a nonlinear Josephson resonator, driven near its double plasma frequency, as a sensitive quantum detector. In this regime the system may develop a bifurcation with two possible stable states; it may be manipulated to force it to the state, correlated with the state of a coupled qubit. In contrast to the Josephson bifurcation amplifier, one of these states has a zero amplitude, which simplifies the task of resolving the two states. Furthermore, the properties of the detector are different from those of the JBA for similar parameters. In particular, the switching curve may be narrower than that for a JBA, that may result in a higher fidelity of the qubit readout.

We thank Michael Wulf, Ralf Dolata and members of the
Cluster of Excellence QUEST  for useful discussions, and Yuli Nazarov for stimulating comments.
This work was partially supported by the EU through the EuroSQIP and
SCOPE project, which acknowledges the financial support of the
Future and Emerging Technologies (FET) programme within the
Seventh Framework Programme for Research of the European
Commission, under FET-Open grant number 218783, by DFG (German Science Foundation) through the Grant ZO124/2-1,
by RFBR under grant No. 09-02-12282-ofi\_m, MES of RF, and the Dynasty foundation (YM).

\end{document}